\documentclass[12pt]{article}
\usepackage{amsfonts}
\usepackage{bbm}
\usepackage{booktabs}
\usepackage{graphicx}
\usepackage{mathrsfs}
\usepackage{color}
\usepackage{verbatim}
\usepackage{cancel}
\usepackage{subfigure}
\usepackage{algpseudocode}
\usepackage{amsmath}
\usepackage{setspace}
\usepackage{geometry}             
\usepackage{amssymb}
\usepackage{amsmath}
\usepackage{epstopdf}
\usepackage{cases}
\usepackage{cite}

\usepackage[hyperindex=true,
pdfstartview=FitH,
bookmarksnumbered=true,
bookmarksopen=true,
citecolor=blue,
linkcolor=blue,
colorlinks=true,
urlcolor=rossoCP3,
unicode]{hyperref}

\definecolor{rossoCP3}{cmyk}{0,.88,.77,.40}

\begin{document}

\title{\bf Ruppeiner geometry and the fluctuation of the RN-AdS black hole in framework of the extensive thermodynamics}
\author{\small Chao Wang${}^{1,2}${}, Shi-Peng Yin${}^{1}${}, Zhen-Ming Xu${}^{1,2,3,4}${}, Bin Wu${}^{1,2,3,4}$\thanks{{\em email}: \href{mailto:binwu@nwu.edu.cn}{binwu@nwu.edu.cn}}{ }, and Wen-Li Yang${}^{1,2,3,4}$
	\vspace{5pt}\\
	\small $^{1}${\it School of Physics, Northwest University, Xi'an 710127, China}\\
	\small $^{2}${\it Institute of Modern Physics, Northwest University, Xi'an 710127, China}\\
	\small $^{3}${\it Shaanxi Key Laboratory for Theoretical Physics Frontiers, Xi'an 710127, China}\\
	\small $^{4}${\it Peng Huanwu Center for Fundamental Theory, Xi'an 710127, China}
}

\date{}
\maketitle

\begin{spacing}{1.2}
	
\begin{abstract}
The Ruppeiner geometry and the fluctuation for a four-dimensional charged AdS black hole in the framework of the extensive thermodynamics have been investigated. By fixing the AdS radius of Visser's construction, the extensive thermodynamics of the RN-AdS black hole is established, and the central charge plays the role of particle number. With the thermodynamic variables' density, we reexamine the phase transition of the black hole, and the result will hold for the black holes under different gravitational models. Moreover, we have studied the Ruppeiner geometry and the black hole's fluctuation, and show that the increase of central charge will reduce the fluctuation probability.
\end{abstract}

\section{Introduction}

By treating the negative cosmological constant $\Lambda=-(d-1)(d-2)/l^2$ in $d$-dimensional spacetime as thermodynamic pressure $P=-\Lambda /8 \pi G_d$ (here $l$ is the AdS radius), extended phase space of the AdS black hole is established, with the sacrifice that reinterpreted the black hole mass as enthalpy. Furthermore, all the thermodynamic variables are related by the Smarr formula \cite{Smarr:1972kt,Caldarelli:1999xj}, instead of the Euler relation, which leads to the result that the black hole thermodynamics lost the extensive characteristics. In standard thermodynamics, the thermodynamic potential is a one-order homogeneous function of the extensive variables \cite{1999Statistical}. For instance, the Euler formula with internal energy $E=T S + y_i x^i$ ($x^i$ is the extensive variables that conjugate to chemical potential $y_i$) is different from the Smarr relation for black holes. Applying the RN-AdS black hole as an example
\begin{equation}
	M=\frac{d-2}{d-3} TS + \Phi Q - \frac{2}{d-3} P V. \nonumber
\end{equation}
$E$, $T$ and $S$, respectively, correspond to the system's internal energy, temperature and entropy. $Q$ is the electric charge, and $\Phi$ is the corresponding potential. The absence of this feature in black hole thermodynamics is quite abrupt since the horizon is an iso-potential surface in the gravitational field. 

There is a turnaround when considering the Anti-de Sitter/Conformal Field Theory correspondence (AdS/CFT) duality \cite{Maldacena:1997re,Gubser:1998bc,Witten:1998qj}. In several works, it has been suggested that the variation of $\Lambda$ corresponds to varying the number of colors $\mathcal{N}$, or the number of degrees of freedom $\mathcal{N}^2$ \cite{Henningson:1998gx,Freedman:1999gp,Myers:2010xs}. For a conformal field theory, $\mathcal{N}^2$ is given by the central charge, which can be expressed with $C=l^{d-2}/G_d$. Here $G_d$ is associated with the Newtonian constant in $d$-dimensional spacetime. Via applying the central charge as a thermodynamic variable, Visser proposed a description of the thermodynamics of boundary field theory, showing that the varying of $\Lambda$ only leads to the variation of $C$ \cite{Visser:2021eqk}. They further derived the holographic Euler equation and the first law as
\begin{align}
	&E = T S + \hat{\Phi} \hat{Q} + \Omega J + \mu \mathrm{d} C, \nonumber\\
	\mathrm{d} E = &T \mathrm{d} S - \mathcal{P} \mathrm{d} \mathcal{V} + \hat{\Phi} \mathrm{d} \hat{Q} + \Omega \mathrm{d} J + \mu \mathrm{d} C, \nonumber
\end{align}
where $E$ is the internal energy of the black hole, $\Omega$, $J$ correspond to the angular velocity and angular momentum. $\hat{Q}$ and $\hat{\Phi}$ are the rescale charge and electric potential, respectively. Moreover, $\mathcal{P}$ is the pressure of the conformal field, and $\mathcal{V}$ is the volume of the boundary. $C$ is the central charge, which plays the same role as the particle number in classical thermodynamics. $\mu$ is the chemical potential conjugated to the central charge that can be deduced from the Euclidean action $A_E$ in the bulk spacetime with a black hole, which is $T A_E =\mu C$. Inspired by this, abundant research about holographic thermodynamics, as well as their thermodynamic properties have been proposed \cite{Rafiee:2021hyj,Cong:2021fnf,Cong:2021jgb}. 

However, there is still no better holographic explanation for the pressure. In this sense, the restricted phase space of the black hole has been proposed by fixing the AdS radius as a constant \cite{Zeyuan:2021uol}, with which the volume term is excluded. It is proved that variables such as entropy are extensive as ordinary thermodynamic systems, and the black hole mass reverts its role of internal energy in this framework. Now, this method has been generalized to asymptotic flat and der Sitter (dS) black holes \cite{Gao:2021xtt,Wang:2021cmz,Zhao:2022dgc}.

Despite the establishment of extensive black hole thermodynamics, there is still a question about its underlying microstructure. The thermodynamic geometry, especially the Ruppeiner metric \cite{PhysRevA.24.488,Ruppeiner:1983zz,Ruppeiner:1995zz}, shows some light on that subject. From the fluctuation theory, Ruppeiner comes out with a line element to measure the distance between the neighboring fluctuation states. Longer distance correspond to a less probability of fluctuation between two neighboring fluctuation states.

Further research reveals that the curvature scalar contains more information. Firstly, all the diagonal metric elements have physical meaning \cite{Ruppeiner:1995zz}, for instance, the heat capacity, which implies that there is correspondence between the phase transition and the singularities of the corresponding curvature scalar. Secondly, the sign of curvature scalar has been provide as an useful tool to test the interaction between the micromolecules of a thermodynamic system: \cite{Ruppeiner:2008kd,Ruppeiner:2010,Sahay:2010wi,Zhang:2015ova,Wei:2017icx,Chaturvedi:2017vgq,Xu:2019gqm,Xu:2020ftx}: $R>0$ means the interaction performance as repulsive domain, while $R<0$ is contrary, and $R=0$ represents there is no interaction. What's more, the curvature scalar was also suggested to be related to the correlation length $\xi$ by $R \sim \kappa \xi^{\bar{d}}$ \cite{Ruppeiner:1995zz,Wei:2019yvs}, there $\kappa$ and $\bar{d}$ represent a dimensionless constant and physical dimensionality of the system.

As discussed above, we are curious about the behavior of Ruppeiner geometry for the RN-AdS black hole in extensive thermodynamics. In this paper, we first show the small/large black hole phase transition \cite{Kubiznak:2012wp,Kastor:2009wy,Dolan:2011xt,Cvetic:2010jb,Wei:2019uqg,Wei:2019yvs} of the black hole with the density, that is obtained from dividing the extensive variables by the central charge, which makes the result hold for AdS black holes in different gravitation models. Then, we reexamined the Ruppeiner geometry. Our calculation shows that the fluctuation probability and line element reduce as the increase of $C$. The black hole is treated as a minor part of an isolated equilibrium system when constructing the thermodynamic geometry with fluctuation theory. Result that as the increase of central charge, the small part will increase, causing a decrease in its fluctuation relative to the entire system. Further, checking the curvature scalar implied that only the interaction in the large black hole with lower temperature act as the repulsive domain. In the end, we get the relative fluctuation of the thermodynamic variables of the black hole. 

The outline of this paper is as follows. In Section \ref{II}, the extensive black hole thermodynamics are briefly introduced, in which the Van der Waals-like phase transition of RN-AdS black holes is further investigated. The Ruppeiner and fluctuation of the black hole are studied in Section \ref{III}. See the conclusions in Section \ref{IIII}. Throughout this paper, we apply the units $\hbar=\kappa_B=c=1$.

\section{Black hole in extensive thermodynamics}\label{II}

In this section, we would like to briefly review the extensive thermodynamics of the RN-AdS black hole, and further study the behavior of variables' density. 

\subsection{Extensive thermodynamics for RN-AdS black hole}
The RN-AdS black hole in four-dimensional spacetime can be described by the action as follow
\begin{equation}
	I =  \frac{1}{2 \kappa} \int_{\mathcal{M}} \mathrm{d}^4 x \sqrt{-g}(\mathcal{R}-2 \Lambda)-\frac{1}{4 \mu_0} \int_{\mathcal{M}} \mathrm{d}^4 x \sqrt{-g}\left(F_{\mu \nu} F^{\mu \nu}\right), \nonumber
\end{equation}
in which $\mathcal{R}$ is the spacetime curvature scalar. $\kappa$ stands for the 4-dimensional Einstein constant, and $\mathcal{M}$ is the spacetime manifold. The electromagnetic field tensor $F_{\mu \nu}$ is
\begin{equation}
	F_{\mu \nu} =\nabla_\mu A_\nu - \nabla_\nu A_\mu, \nonumber
\end{equation}
where $A_\mu= \left( -\Phi(r),0,0,0 \right)$ is the vector potential. To get the solution of spacetime, one always assumes the line element which has the form of
\begin{equation}
	\mathrm{d} s^2 = - f(r) \mathrm{d} t^2 + \frac{\mathrm{d} r^2}{f(r)} +r^2 \mathrm{d} \theta^2 + r^2 \sin^2 \theta \mathrm{d} \phi^2, \nonumber
\end{equation}
and the metric function $f(r)$ can be solved as
\begin{equation}
	f(r)=1- \frac{2 G M}{r} + \frac{G Q^2}{r^2} +\frac{r^2}{l^2}, \quad \Phi(r) = \frac{Q}{r}. \nonumber
\end{equation}

Now let's turn our attention to the thermodynamic properties of the black hole. Regarding the central charge $C=l^2/G$ of the Conformal Field theory as a thermodynamic variable of the black hole, it plays a role similar to that of the quantity of matter. According to the AdS/CFT correspondence, the charge and electric potential should be rescaled as \cite{Visser:2021eqk,Cong:2021fnf}
\begin{equation}
	\hat{Q}= \frac{Q l}{\sqrt{G}}, \quad \hat{\Phi}=\frac{\Phi(r_+) \sqrt{G}}{l}= \frac{Q \sqrt{G}}{r_+ l}, \nonumber
\end{equation}
and the black hole mass and entropy are \cite{Zeyuan:2021uol}
\begin{equation}
	M(S, \hat{Q}, C)=\frac{S^2 +\pi S C +\pi^2 \hat{Q}^2}{2\pi^{3/2} l (S C)^{1/2}}, 
	\quad S=\frac{\pi r_+^2}{G}.  \label{M}
\end{equation}
Therefore the first law and equation of state (EoS) of the RN-AdS black hole in extensive thermodynamics framework has the form of
\begin{align}
	\mathrm{d} M &=T \mathrm{d} S+\hat{\Phi} \mathrm{d} \hat{Q}+\mu \mathrm{d} C, \label{fl1}\\
	M &=T S+\hat{\Phi}  \hat{Q}+\mu C, \label{EOS}
\end{align}
where $\mu$ is the chemical potential conjugate to $C$. The equation Eq.(\ref{EOS}) indicates that the black hole mass is a first-order homogeneous function of the quantities ($S, \hat{Q}, C$). According to the first law in Eq.(\ref{fl1}), the thermodynamic variables of the black hole can be deduced easily
\begin{align}
	T &=\left(\frac{\partial M}{\partial S}\right)_{\hat{Q}, C}=\frac{3 S^{2}+\pi S C-\pi^{2} \hat{Q}^{2}}{4 \pi^{3 / 2} l S(S C)^{1 / 2}}, \label{T}\\
	\hat{\Phi} &=\left(\frac{\partial M}{\partial \hat{Q}}\right)_{S, C}=\left(\frac{\pi}{S C}\right)^{1 / 2} \frac{\hat{Q}}{l}, \label{Phi}\\
	\mu &=\left(\frac{\partial M}{\partial C}\right)_{S, \hat{Q}}=-\frac{S^{2}-\pi S C+\pi^{2} \hat{Q}^{2}}{4 \pi^{3 / 2} l C(S C)^{1 / 2}}, \label{mu}
\end{align}
which are intensive variables, which will not be rescaled when $(S, \hat{Q}, C)\rightarrow (\lambda S,\lambda \hat{Q},\lambda C)$, where $S$ and $\hat{Q}$ are proportional to $C$ \cite{Zeyuan:2021uol}. Moreover, the dimension of these intensity quantities are
\begin{equation}
	[ T ] = [ \hat{\Phi} ] =[\mu]=\frac{1}{ [ l ] }, \label{scala}
\end{equation}
which will be applied for the analysis of the black hole thermodynamic behavior in the next subsection.

\subsection{Thermodynamic behaviors}

In this subsection, we will revisit the thermodynamic behavior of the black hole in extensive thermodynamics. As the Van der Waals fluid \cite{1999Statistical}, the phase structure in previous studies was obtained in the reduced parameter space, with the critical point denoted by the black bole charge. For example, the reduced entropy was marked as $\tilde{S} \sim S/Q^2$. Thus, the physical meaning of the reduced parameter is ambiguous. In this paper, we mainly explore the variation of the thermodynamic quantity density in the view of the role of particle number of the central charge, which are
\begin{equation}
	s=S/C, \quad q=\hat{Q}/C, \quad t=Tl. \nonumber
\end{equation}
Here $t$ is the dimensionless temperature with the scale analysis Eq.(\ref{scala}). The density variation takes into account the change in the extensive variables and the central charge, so that the results hold for black holes under different gravitational theories. And the EoS comes to
\begin{equation}
	t=\frac{3s^2+\pi s - \pi^2 q^2}{4 (\pi s)^{3/2} }, \label{eos}
\end{equation}
The physical allowed black hole $(t>0)$ satisfied
\begin{equation}
	s \geq \frac{\pi }{6} \left( \sqrt{12 q^2 +1}-1 \right) \nonumber
\end{equation}

For RN-AdS black hole, there exist a small/large black hole phase transition with a critical point satisfied 
\begin{equation}
\left(\frac{\partial t}{\partial s}\right)_{q}=\left(\frac{\partial^{2} t}{\partial s^{2}}\right)_{q}=0, \label{cp}
\end{equation}
which can be denoted as
\begin{equation}
	s_c = \frac{\pi}{6}, \quad q_c = \frac{1}{6}, \quad t_c = \frac{\sqrt{6}}{3 \pi}, \nonumber
\end{equation}
with universal value.

\begin{figure}[!h]
	\centering
     	\includegraphics[width=7cm]{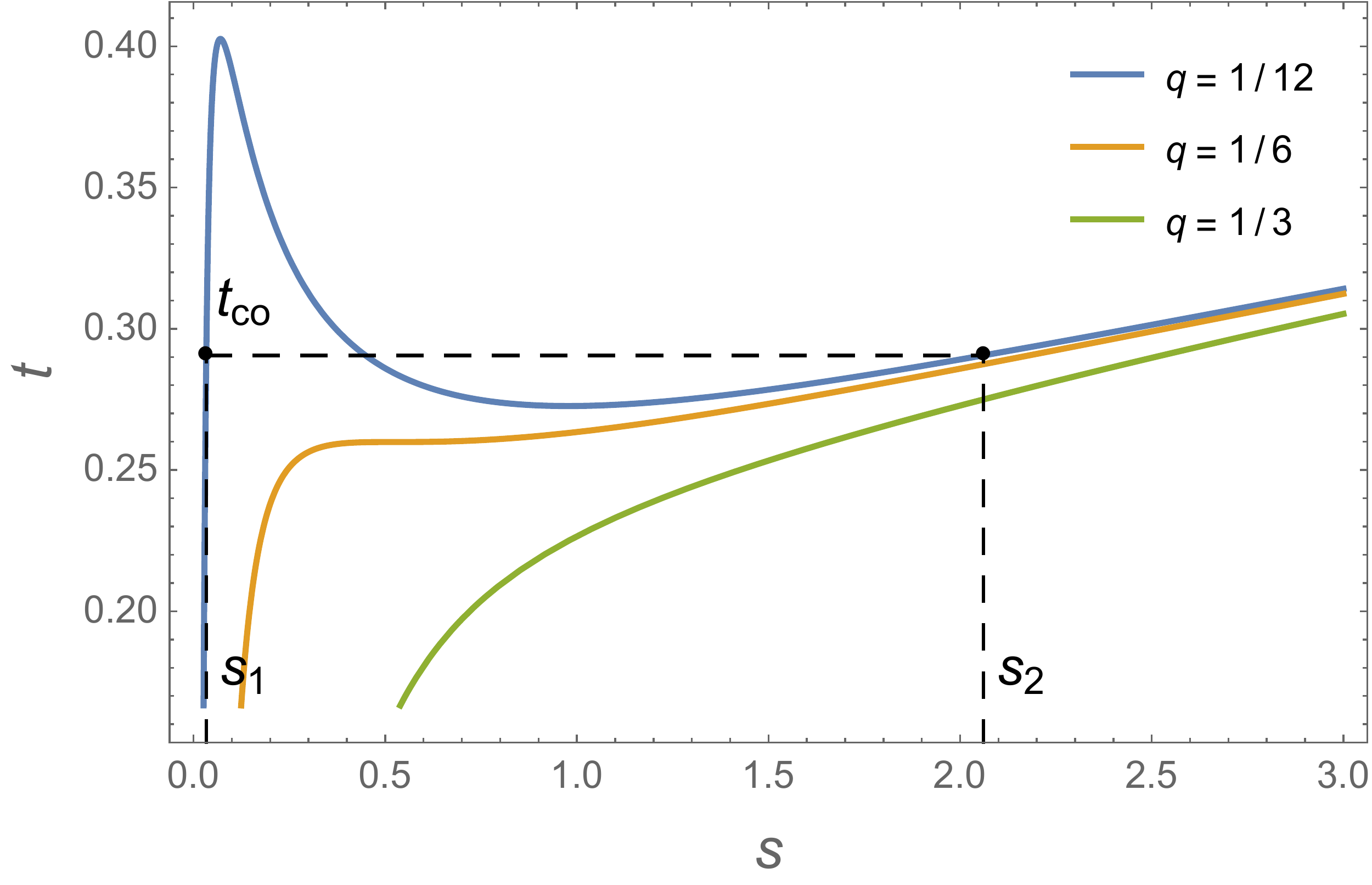}
\caption{$t-s$ curve with $q=1/12$, $1/6$ and $1/3$ from top to bottom. The black dashed lines form the Maxwell equal area law with $t_{\mathrm{co}}$ and $s_{1,2}$ denote the coexistence temperature and saturated small/large black hole. }\label{tsq}
\end{figure}

With the dimensionless EoS in Eq.(\ref{eos}), we depict the iso-$q$ curve with charge density $t=1/12$, $1/6$ and $1/3$ from top to bottom in Fig.\ref{tsq}.  
As shown in the figure, the iso-charge process exhibits oscillatory behavior that resembles the van der Waals phase transition at a value of $q$ less than the critical value, indicating there is a one-order phase transition being experienced by the black hole. With the Maxwell equal area law established by the black dashed line, the phase transition temperature, as well as the corresponding entropy, can be deduced. In this case, as the temperature of a small black hole increases to coexistence value $t_{\mathrm{co}}$, its entropy is expected to jump from $s_1$ to $s_2$, followed by the completion of the transition from a stable small black hole to a stable large black hole. As the black hole charge arrives at the critical value, the two local extrema points merge into an inflection point, at which the system undergoes a second-order phase transition. This time the black hole on the orange curve is thermal stable. Keep increasing the charge and the black hole will be in a supercritical state. The above analysis demonstrates that the RN-AdS black hole with more electric charge has a greater chance to stay in the thermal stable phase.

Let us delve into the first-order phase transition of a black hole, whose core is the  Maxwell equal area law that satisfied
\begin{equation}
	t_{\mathrm{co}}\left(s_{2}-s_{1}\right)=\int_{s_{1}}^{s_{2}} t(s) \mathrm{d} s, \quad t_{\mathrm{co}}=t\left(s_{1}\right)=t\left(s_{2}\right), \quad t_{\mathrm{co}}=\frac{t\left(s_{1}\right)+t\left(s_{2}\right)}{2}, \nonumber
\end{equation}
and the saturated black hole's temperature and entropy are
\begin{align}
	t_{\mathrm{co}}&=  \frac{ \sqrt{ 1-2 q } }{ \pi }   \label{tco1}\\
	s_{1,2}&= \frac{\pi (\sqrt{1-2q} \pm \sqrt{1-6q})^2}{4}. \label{cos}
\end{align}
By the series expansion of $\Delta s = s_2 - s_1$ near the critical temperature, we can obtain the behavior of the black hole at the critical point, which is
\begin{align}
	\Delta s = s_2 - s_1=\frac{2 \sqrt{6 }  \pi }{3} \left(\frac{t-t_c}{t_c}\right)^{\frac{1}{2}} + \frac{5 \sqrt{6 }  \pi }{6} \left(\frac{t-t_c}{t_c}\right)^{\frac{3}{2}}  + \mathcal{O}   \left[ \frac{t-t_c}{t_c} \right]^{\frac{5}{2}}, \nonumber
\end{align}
where $t_c=\frac{\sqrt{6}}{3 \pi}$ is the temperature of critical point. The series expansion implied that the critical exponent of the RN-AdS black hole takes the universal value of $1/2$ which is the same as the Van der Waals fluid.

\begin{figure}[!h]
	\centering
	\includegraphics[width=8cm]{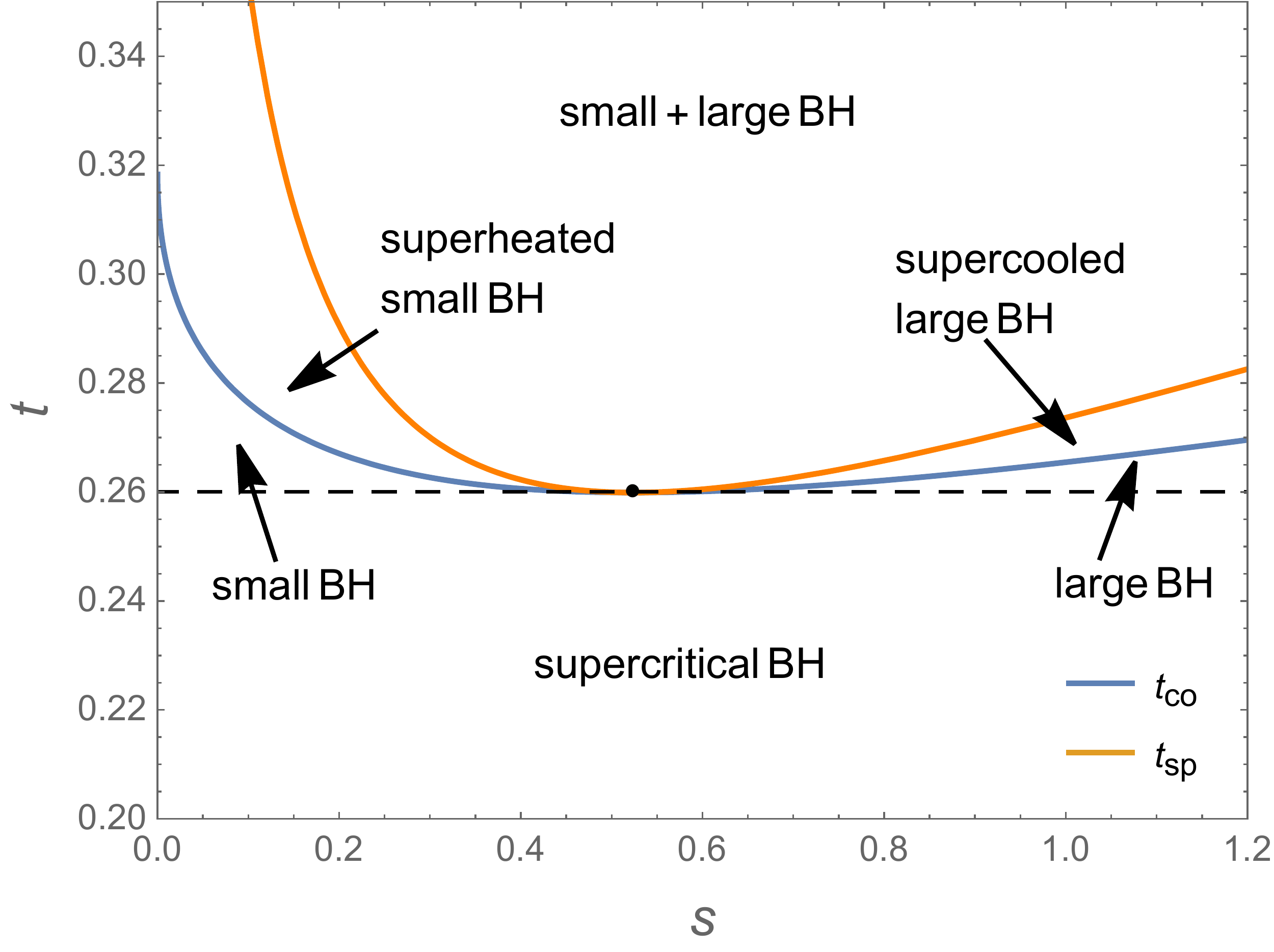}
	\caption{Phase structure of the RN-AdS black hole. The blue and orange curves correspond to the coexistence curve and the spinodal curve, respectively. The black dashed line matches the critical temperature, while the black dot is the critical point of the phase transition.}\label{tsphase}
\end{figure}

The local extremum point of the first-order phase transition as the demarcation between metastable and unstable states is also of interest, which fulfills $(\partial_{s} t)_{q,c}=0$, and one can get the spinodal curve as
\begin{equation}
	t_\mathrm{sp}= \frac{\sqrt{6} \left(1 \pm \sqrt{1-36 q^2} -12 q^2\right)}{2 \pi \left(1\pm \sqrt{1-36q^2}\right)^{3/2}}, \quad
	s_{3,4}= \frac{\pi \left(1 \pm \sqrt{1-36 q^2}\right)}{6}. \label{tsp1}
\end{equation}
Substituting $s_{1,2}$ and $s_{3,4}$ into the coexistence curve $t_\mathrm{co}$ and spinodal curve $t_{\mathrm{sp}}$, the phase structure of the black hole in iso-charge process is represented in Fig.\ref{tsphase}.
The blue and orange curves correspond to the coexistence curve and the spinodal curve, respectively. The black dashed line matches the critical temperature, while the black solid dots are the critical point of the phase transition. In the iso-charge process, it is observed that the two curves divide the phase space into several states, and the black holes with lower temperature are more likely to be left in the supercritical state.

\section{Thermodynamic geometry of the black hole}\label{III}

In this section, we are going to briefly review the Ruppeiner geometry, which provide a possible way to explore the microstructure of black holes.

Consider an equilibrium isolated thermodynamic system with total entropy $S$, which is divided into two subsystems of different sizes. The entropy of the small size is denoted as $S_B$, and the large one $S_E$ can be regarded as the thermal bath. The total entropy is additive and not conserved, while other extensive variables are additive and coversed. Further, we require that $S_B \ll S_E \sim S$. The total entropy described by independent thermodynamic parameters $x^\mu$ have the form of 
\begin{equation}
	S(x^\mu)=S_B(x^\mu)+S_E(x^\mu), x^\mu=(x^1,x^2,\cdots),  \nonumber
\end{equation}
here $x^\mu$ represent the thermodynamic variables. Expand the total entropy at the vicinity of the local maximum ($x^\mu = x^\mu_0$), and we get
\begin{align}
	\Delta S = S- S_0=& \left.\frac{\partial S_B}{\partial x^\mu}\right|_{x_0^\mu} \Delta x_B^\mu+\left.\frac{\partial S_E}{\partial x^\mu}\right|_{x_0^\mu} \Delta x_E^\mu \nonumber\\
	&+\left.\frac{1}{2} \frac{\partial^2 S_B}{\partial x^\mu \partial x^\nu}\right|_{x_0^\mu} \Delta x_B^\mu \Delta x_B^\nu+\left.\frac{1}{2} \frac{\partial^2 S_E}{\partial x^\mu \partial x^\nu}\right|_{x_0^\mu} \Delta x_E^\mu \Delta x_E^\nu+\cdots, \label{deltas}
\end{align}
with $S_0$ as the entropy at the local maximum point. For the equilibrium isolated system, the stability conditions are
\begin{equation}
	\delta S=0, \quad \delta^2 S < 0.
\end{equation}
Via the Gibbs-Duhem relation \cite{Zeyuan:2021uol} and the first law Eq.(\ref{fl1}), we obtain
\begin{equation}
	\delta^2 S = - \sum_i \frac{1}{T_i} \left( \delta T_i \delta s_i +\delta {\hat{\Phi}}_i \delta q_i   \right) C_i, \nonumber
\end{equation}
where $i$ represents the corresponding segments of the entire system. It is obvious that $\delta^2S$ is independent of the perturbation of $C$.. Therefore, the number of particles cannot be used as thermodynamic coordinates. Since the thermal bath gets the same order as that of the whole system, so the second derivatives with respect to the thermodynamic variables are much smaller than that of $S_B$ and thus can be ignored. Therefore the $\Delta S$ arrives at
\begin{equation}
	\Delta S =  \left.\approx \frac{1}{2} \frac{\partial^2 S_B}{\partial x^\mu \partial x^\nu}\right|_{x_0^\mu} \Delta x_B^\mu \Delta x_B^\nu,
\end{equation}
which implied that the fluctuations of the whole isolated system are mainly provided by small sub-systems.
 
In view of the fluctuation theory, the probability of a system being in a certain fluctuation state \cite{PhysRevA.24.488,Wei:2019yvs} is
\begin{equation}
	P(x^\mu)=W e^{-\frac{1}{2} \Delta l^2}, \label{pro}
\end{equation}
where $x^\mu$ are the independent thermodynamic variables. $W$ is a normalization constant. The thermodynamic line element measures the distance between two neighboring fluctuation states denoted by $\Delta l^2$ has the form of
\begin{align}
	\Delta l^2 &=  -g_{\mu \nu} \Delta x^\mu \Delta x^\nu, \nonumber\\
	g_{\mu \nu}&= \frac{\partial^2 S}{\partial x^\mu \partial x^\nu}. \nonumber
\end{align}

Now we consider the RN-AdS black hole in extensive thermodynamics. Choose temperature $t$ and charge $\hat{Q}$ of the black hole as the thermodynamic coordinates to reveal its underlying microstructure. The line element $\Delta l^2$ is given by \cite{Wei:2019yvs}
\begin{align}
	\Delta l^2 &= g_{tt} \Delta t^2 +g_{ {\hat{Q}} {\hat{Q}} } \Delta \hat{Q}^2, \nonumber\\
	  &=\frac{1}{t} \left(\frac{\partial S}{\partial t}\right)_{ \hat{Q} } \Delta t^2 + \frac{1}{t} \left(\frac{\partial \hat{\Phi}}{\partial \hat{Q}}\right)_t \Delta \hat{Q}^2 \sim C. \label{line}
\end{align}
Combined with the equation Eq.(\ref{pro}), it can be seen that increased $C$ leads to decreased probability of a fluctuation of the thermodynamic system.  Moreover, the relative fluctuation of entropy and charge are \cite{1999Statistical}
\begin{align}
	\frac{ \sqrt{ \overline{(\Delta t)^2} } }{t} &=  \sqrt{ \frac{3 S^2 -\pi S C +3 \pi^2 \hat{Q}^2}{2S(3 S^2 + \pi S C - \pi^2 \hat{Q}^2) } } \sim \frac{1}{\sqrt{C}}, \label{fs}\\
	\frac{ \sqrt{ \overline{(\Delta \hat{Q})^2} } } {\hat{Q}} &= \sqrt{\frac{3 S^2 + \pi S C - \pi^2 \hat{Q}^2 }{ 4 \pi^2 S \hat{Q}^2 }  } \sim \frac{1}{\sqrt{C}}, \label{fq}
\end{align}
which indicates that the relative fluctuation of the RN-AdS black hole thermodynamic variables decreases monotonically with the increase of $C$. Fluctuations of other thermal parameters follow the same pattern. 

\begin{figure}[!h]
	\centering
		\quad
	\subfigure[q=1/12]{
		\includegraphics[width=4.5cm]{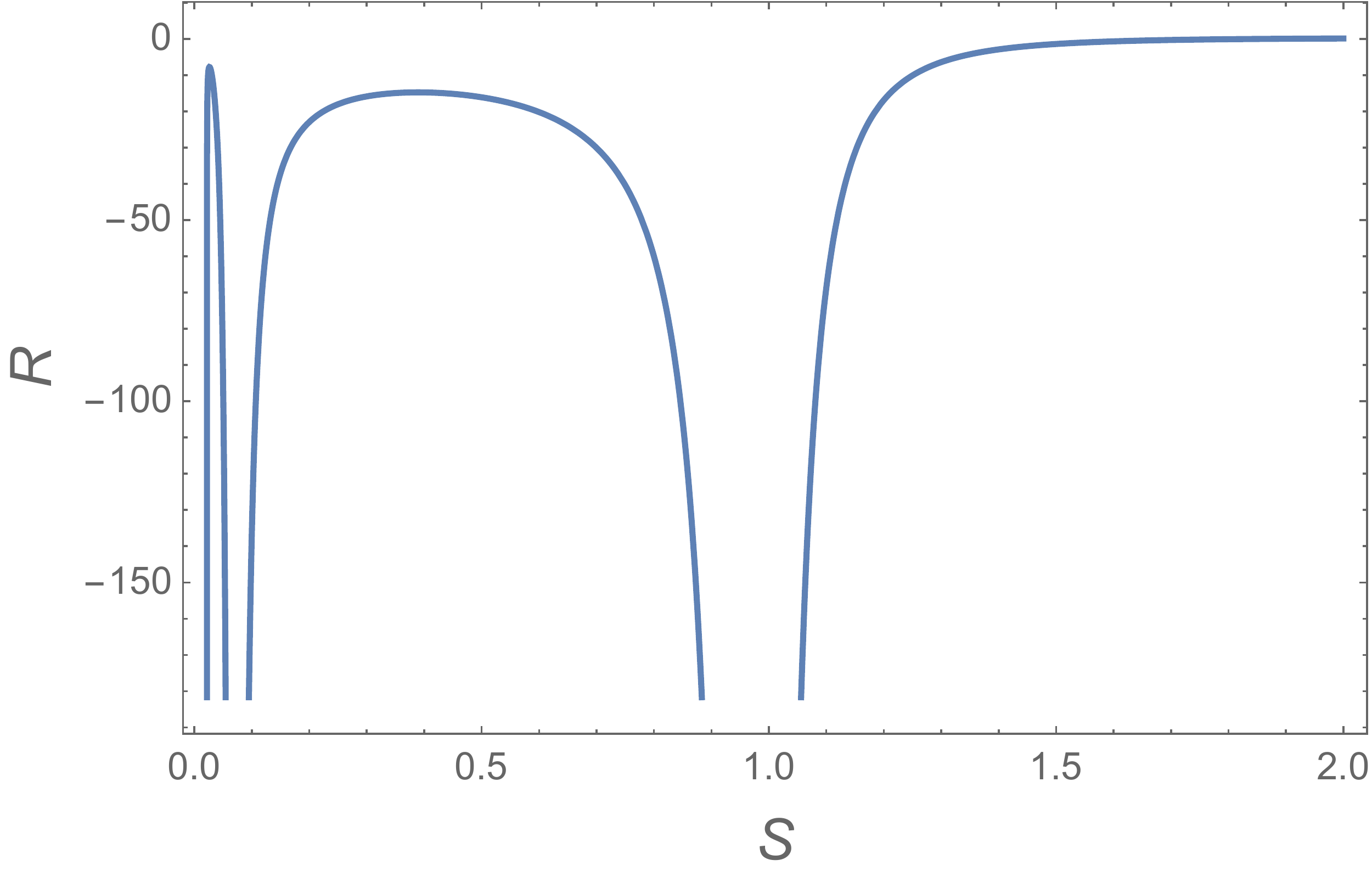}
	}
	\subfigure[q=1/6]{
		\includegraphics[width=4.5cm]{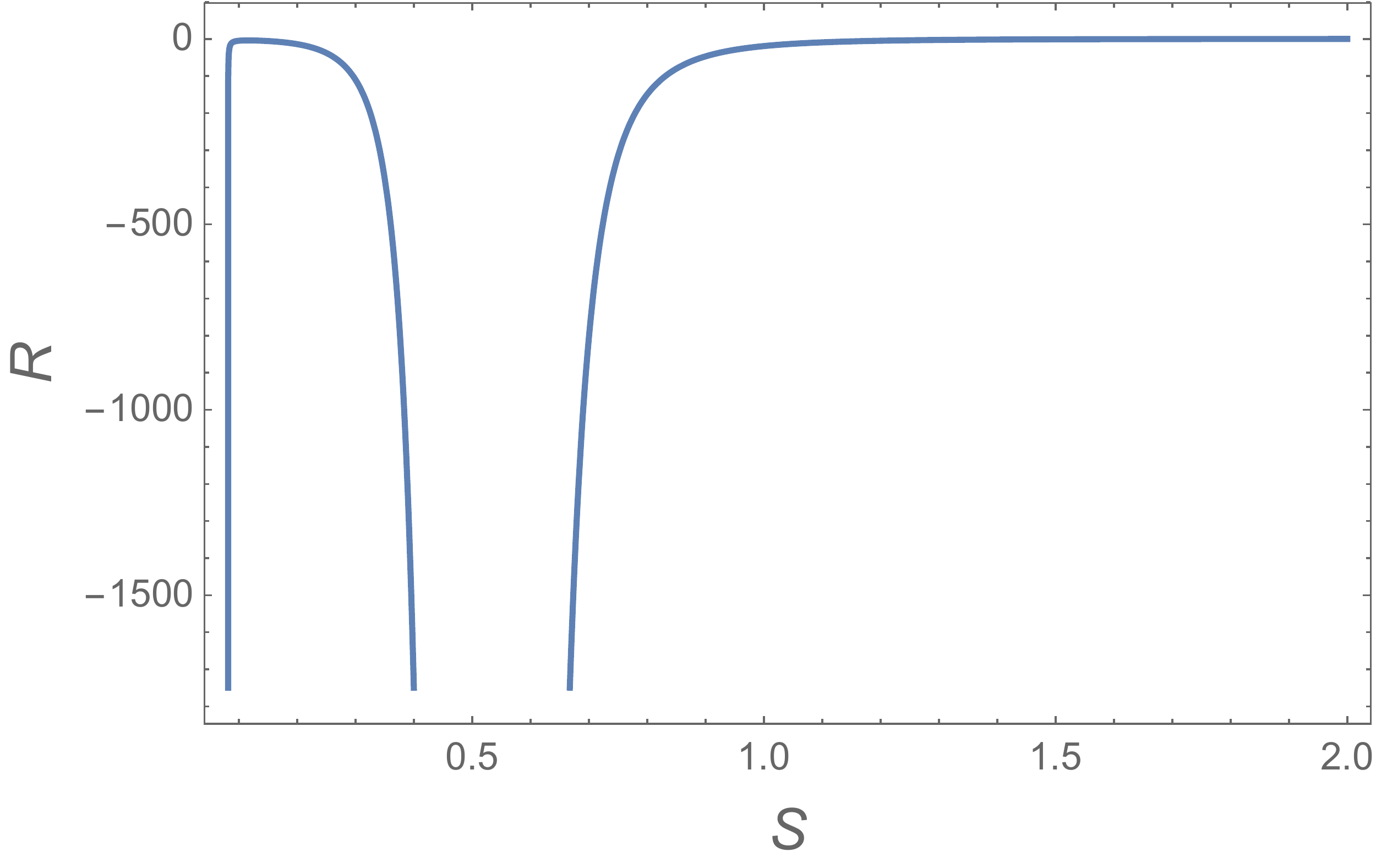}
	}
	\quad
	\subfigure[q=1/3]{
		\includegraphics[width=4.5cm]{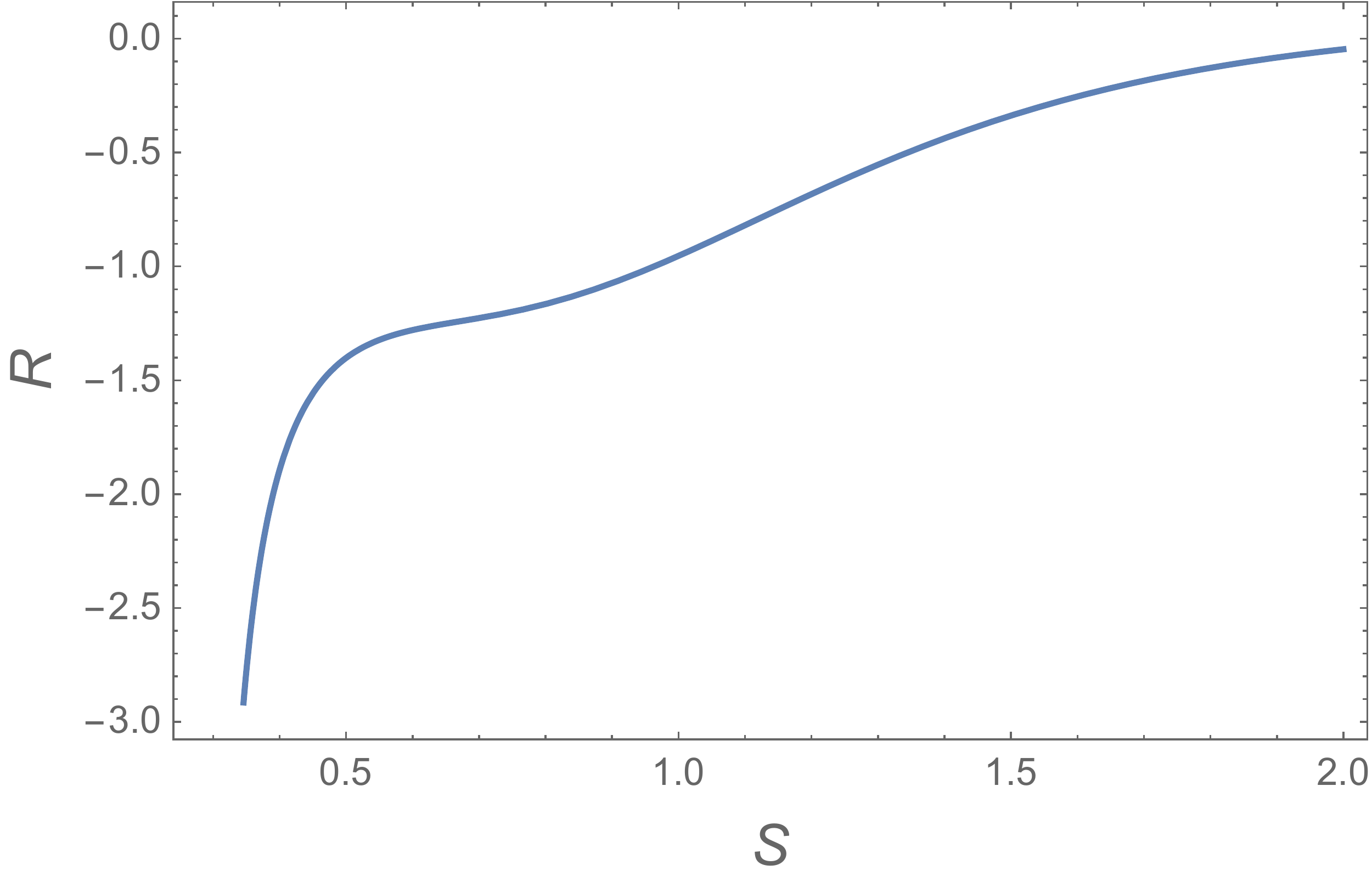}
	}
	\caption{The behavior of the thermodynamic curvature scalar with $\hat{Q}=1/12$, $1/6$ and $1/3$. Here we set $C=1$.}\label{Rsq}
\end{figure}

Another important topic is the thermodynamic curvature scalar. First, the divergence point of the curvature scalar corresponds to the phase transition point of the system, and second, the sign of it is related to the interactions. The curvature scalar of the RN-AdS black hole in extensive thermodynamics is 
\begin{align}
	R=&\frac{6 \pi^2 \hat{Q}^2 S \left( 5 \pi^2 \hat{Q}^2 + 9 S^2 \right) + \pi^2 C^2 S \left( 9S^2 - 2\pi ^2 \hat{Q}^2 \right) }{ \left(\pi^2 \hat{Q}^2 - \pi S C - 3 S^2 \right) \left( 3 \pi^2 \hat{Q}^2 - \pi S C +3 S^2\right)^2  } \nonumber\\
	&+ \frac{   \pi C \left( \pi^4 \hat{Q}^4 -27 \pi^2 \hat{Q}^2 S^2+ \pi^2 S^2 C^2 -18 S^4 \right)}{  \left(\pi^2 \hat{Q}^2 - \pi S C - 3 S^2 \right) \left( 3 \pi^2 \hat{Q}^2 - \pi S C +3 S^2\right)^2  }. \label{R}
\end{align}
To describe the thermodynamic properties of the black hole clearly, we show the behavior of curvature scalar $R$ with $\hat{Q}=1/12$, $1/6$ and $1/3$ in Fig.\ref{Rsq}. Here we set $C=1$.
We can see that $R$ starts with a divergent piont which corresponds to $t=0$. When the small/large black hole occur, the curvature scalar has two divergent points determined by the condition  $(\partial_S T)_{ \hat{Q} }=0$, which corresponds to the spinodal curve. As the increase of charge, these two divergent points merge into one, which indicates that the phase transition shifts from the first-order to the second-order. Keep increasing the charge, the divergent point will disappear, and the black hole are in the supercritical phase. 

\begin{figure}[!h]
	\centering
	\includegraphics[width=7cm]{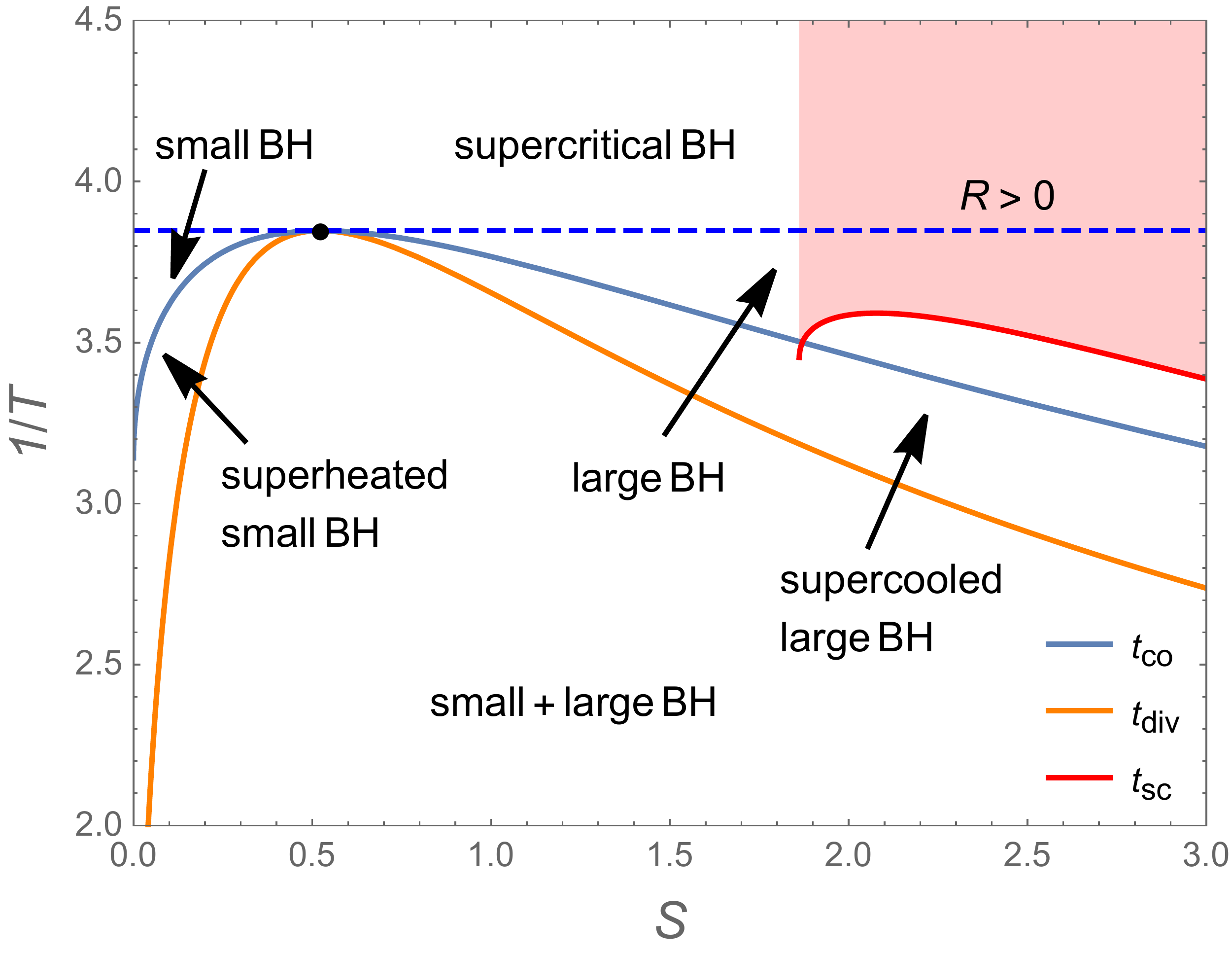}	
	\caption{The characteristic curve of curvature scalar in coordinate space $\{T, S\}$. The orange and red solid curves are the divergence curve and the sign-changing curve of the curvature scalar, respectively. The gray solid curve is the coexistence curve in Eq.(\ref{tco1}). The blue dashed line in the figure corresponds to the critical point temperature.}\label{charac}
\end{figure}

As discussed above, checking the characteristic curve including the sign-changing curve $T_\mathrm{sc}$ and divergent curve $T_\mathrm{div}$ of curvature scalar would be helpful for the study of the black hole phase transition and underlying microstructure. With Eq.(\ref{T}), Eq.(\ref{tco1}) and Eq.(\ref{R}), we plot the characteristic curve in Fig.\ref{charac}.
The gray solid curve in the figure corresponds to the coexistence curve which is based on Eq.(\ref{tco1}) and Eq.(\ref{cos}). The extreme point of $T_\mathrm{co}$ divides the coexistence curve into saturated small phase and saturated large phase. The orange and red curves are the divergence curve and sign-changing curves, respectively. These characteristic curves distinguish the states in which the corresponding black holes are located. More importantly, the sign-changing curve points out the information about the interaction inside the black hole. Positive curvature scalars are found in the region above $T_\mathrm{sc}$ that is painted in red, the interactions within the corresponding black holes are dominated by repulsive. Meanwhile, the interactions in the other regions behave as an attractive domain. 

Finally, we will further analyze the property of $R$ converging to the critical point along the coexistence curve. Combining Eq.(\ref{tco1}), Eq.(\ref{cos}) and Eq.(\ref{R}), the critical behavior of $R$ can be described as
\begin{align}
	R(\mathrm{SSBH})&= -\frac{1}{10 \pi} \left( \frac{ t - t_c }{t_c}  \right)^{-2} - \frac{\sqrt{6}}{32 \pi } \left( \frac{ t - t_c }{t_c}  \right)^{-3/2} + \mathcal{O} \left[\frac{t-t_c}{t_c}\right]^{-1}, \\
	R(\mathrm{SLBH})&= -\frac{1}{10 \pi} \left( \frac{ t - t_c }{t_c}  \right)^{-2} + \frac{\sqrt{6}}{32 \pi } \left( \frac{ t - t_c }{t_c}  \right)^{-3/2}  + \mathcal{O} \left[\frac{t-t_c}{t_c}\right]^{-1}, \nonumber
\end{align}
where SSBH and SLBH correspond to the saturated small and large black holes, respectively. It is noticed that the critical exponent is equal to $2$, which is consistent with that of the Van der Waals fluid.

\section{Conclusion}\label{IIII}

In this paper, we have studied the Ruppeiner geometry and fluctuation of a four-dimensional charged AdS black hole in the extensive thermodynamics. By fixing the AdS radius of Visser's construction \cite{Visser:2021eqk}, the extensive thermodynamics of RN-AdS black hole is established \cite{Gao:2021xtt}, with correct Euler relation and Gibbs-Duhem equation. The central charge $C=l^2/G$ was introduced as a thermodynamic variable, which plays the same role as particle number. In this framework, with the density of thermodynamic variables, we reexamined the small/large phase transition of the RN-AdS black hole. Since the density takes into account the variation of thermodynamic quantities and central charge, our results will hold for RN-AdS black holes under different gravitational theories. We further examine the critical behavior, and obtain a critical exponent $2$ that is the same as Van der Waals fluid.

Then we reviewed the Ruppeiner geometry and studied the microstructure of the black hole. When we get to the Ruppeiner geometry,  finding that the variation of particle number has no influence on the fluctuation of the equilibrium isolate system. In this way, the thermodynamic line element and relative fluctuation of the RN-AdS black hole were constructed. They are all proportional to $1/C$. In fluctuation theory, by dividing the isolated equilibrium system into two sub-systems with different sizes, the fluctuation of the whole system is approximately equal to the smaller one noted as $S_B$. In this sense, the increase of $C$ corresponds to a larger $S_B$, which leads to the reductive of the smaller sub-system fluctuation relative to the whole system. Besides, the interaction has also been studied with the curvature scalar. Our results indicate that only the interaction inside the low-temperature large black hole is a repulsive domain. In the end, we examine the critical behavior, which owns the same value of critical exponent as the gas/liquid phase transition.

As for future work, we will continue to study the thermodynamic macroscopic behaviors of black holes and its differences from classical thermodynamics deeply, and further explore the relationship between thermodynamic geometry and fluctuation.

\section*{Acknowledgment}
This work is supported by the financial supports from the National Natural Science Foundation of China (Grant Nos.12275216, 12047502, 12105222), the China Postdoctoral Science Foundation (Grant Nos.2017M623219, 2020M673460). 

\end{spacing}

\providecommand{\href}[2]{#2}\begingroup
\footnotesize\itemsep=0pt
\providecommand{\eprint}[2][]{\href{http://arxiv.org/abs/#2}{arXiv:#2}}
\providecommand{\doi}[2]{\href{http://dx.doi.org/#2}{#1}}

\endgroup

\end{document}